\documentclass[ 12pt , a4paper  ]{article}
\usepackage[utf8]{inputenc}
\usepackage[english]{babel} 
\usepackage[T1]{fontenc}

\usepackage{amsmath}
\usepackage{graphicx}
\usepackage[labelfont=bf]{caption}
\usepackage[colorlinks=true , linkcolor=blue , citecolor=black]{hyperref}

\usepackage{authblk}

\usepackage{geometry} 
\title{    
Unused power surge compromises\\
U.S. road vehicles sustainability}

\author[1,2]{Hugues Perraut}
\author[1]{Joseph Le Bihan}
\author[1]{Eric Herbert}
\author[1,*]{Petros Chatzimpiros}

\affil[1]{Laboratoire Interdisciplinaire des Energies de Demain (LIED), UMR 8236, Université Paris Cité, 75013 Paris, France}
\affil[2]{Université Paris-Saclay, ENS Paris-Saclay, DER de Physique, 91190 Gif-sur-Yvette, France}
\affil[*]{Corresponding author, petros.chatzimpiros@u-paris.fr}

\date{\today}

\begin{document}

\maketitle 

\section*{Abstract}

Material and energy flows underpin sociotechnical metabolism. However, despite growing sustainability concerns over expanding material stocks and declining stock productivity, the link between material use and energy consumption remains poorly understood. This gap reflects a limited distinction between structures and the activity they enable, and the lack of quantification of the installed power of energy consuming structures. Here we reconstruct the long-term growth dynamics of U.S. road vehicles, distinguishing professional and consumer assets. We show that installed power, mass, and fuel energy use follow divergent patterns within and across vehicle categories. By introducing the usage factor as a metric linking structure to activity, we quantify decoupling mechanisms such as engine oversizing and fleet redundancy, which drive up material immobilization. As electrification requires large-scale fleet replacement, our findings highlight that avoiding power oversized vehicles could reduce material demand, emphasizing the need to account for structure-activity decoupling in energy transition policies.

\section{Introduction}

Material and energy flows underpin the metabolism of sociotechnical systems, shaping their organization, environmental impacts, and sustainability. Material and energy flow analysis (MEFA) has been developed as a quantitative framework to characterize sociotechnical system metabolism and dynamics, with applications ranging from individual micro-level processes to macroscopic system description\,\cite{haberl2019, Krausmann2017}. Analyzing these structural dynamics is critical for sustainability, as accelerating global resource extraction poses socioenvironmental challenges\,\cite{haberl2019,steffen2015}. 
So far, research has illuminated dynamics of total volumes for specific substances and energy carriers, material-energy intensities, society metabolic rates and stock productivity, including the relation of GDP or energy consumption to total in-stock mass\,\cite{Ayres2002, Schandl2024,unitednationsprogram2011}. 
At this macroscopic scale, evidence shows that stock productivity is declining, meaning modern societies involve increasingly larger physical structures to deliver activity\,\cite{Krausmann2017}. 

Nonetheless, applying MEFA to entire socioeconomic metabolism risks overlooking sector-specific structural dynamics or divergent patterns between professional and consumer assets, which may arise from distinct structure–activity relationships across contexts. The literature rarely addresses the distinction between structure and activity in sociotechnical systems\,\cite{Krausmann2017}, despite its critical and physically meaningful implications\,\cite{chen2015}. Structures, produced by industry from materials and energy, are subsequently deployed across sectors such as agriculture, transport, and buildings, where they enable activity -- the use of structures. Comparatively to structures, activity primarily involves energy use, while its material use is marginal\,\cite{Krausmann2017}.  
\newline
Accordingly, energy-to-material use intensity may vary across sectors and between professional and consumer assets, depending on the structure–activity relationship. Sectors where structures grow faster than activity may show rising material consumption per unit energy conversion and vice-versa, highlighting potentially diverging patterns across sectors.

However, the relationship between structure and activity is not straightforward and remains largely unexplored. Structures enable activity by converting energy, but actual energy conversion typically represents only a fraction of a structure’s conversion capacity -- i.e., its installed power in watts. This gap arises from how long and how intensively converters operate, and reflects the degree of oversizing of structures relative to the activity they generate.
While installed power is a well-established metric for production systems -- such as electricity power plants -- it remains completely unqualified for consumption structures. MEFA has never incorporated installed power, leaving the underlying connections with material use and energy consumption largely elusive.

Here, we introduce the installed power of energy consuming structures (IPECS) -- defined as the sum of the nominal power of individual converters in watts -- as a relevant metric to complement MEFA's characterization of sociotechnical systems. We apply the concept to the U.S. road transportation sector and explore the long-term dynamics of structure-to-activity patterns in vehicle fleets. In this context, the IPECS of the road transportation sector is the total motor power of the vehicles.

Based on best available data, we distinguish cars, trailer trucks, and buses. For each vehicle category, we reconstruct the IPECS and  vehicles mass, then compare these to actual fuel consumption adjusted for engine conversion efficiency. Actual fuel consumption is derived from official statistics and is factorized between average effective motor power and duration of use, based on person-kilometer travel data and driving speed surveys. We define the usage factor for each fleet as the ratio of actual fuel consumption to the theoretical energy consumption if installed power operated continuously at full capacity. Using a Kaya-like framework\,\cite{Kaya}, we break the usage factor into two structural dimensions: the load factor (the fraction of total motor power effectively used in average during operation) and the utilization rate (the fraction of time the structure is active).

We find a clear decoupling between structure and activity for cars and trucks, where growth in total motor power and mass has outpaced fuel consumption, which plateaued over the last two decades. This decoupling reflects declining activity per unit structure, though the underlying mechanisms differ. For cars, a decreasing usage factor over a century resulting from increasing engine oversizing (drop in the load factor) against a remarkable stability in the utilization rate. For trucks, the usage factor remained stable until 2000 as opposing trends balanced out: a declining utilization rate was offset by a rising load factor (e.g. higher engine use intensity from increased speed limits). After 2000, the usage factor declined as the utilization rate continued to fall (driven by fleet redundancy inducing reduced traveled distances) while the load factor (i.e. engine use intensity) stabilized.
For buses, the utilization rate and load factor have remained fairly stable over the period, suggesting close coevolution between structure and activity. 
The decoupling between structure and activity challenges sustainability in general, and electrification policies in particular, by increasing material immobilization. Without targeted interventions to regulate IPECS and vehicle mass, the electrification transition risks to provoke an avoidable immobilization of materials in structurally oversized and sparsely utilized vehicles.

\section{Results}

\subsection{Structure-activity patterns and usage factor of road vehicles}
\begin{figure}[h!]
    \centering
\includegraphics[width=\textwidth]{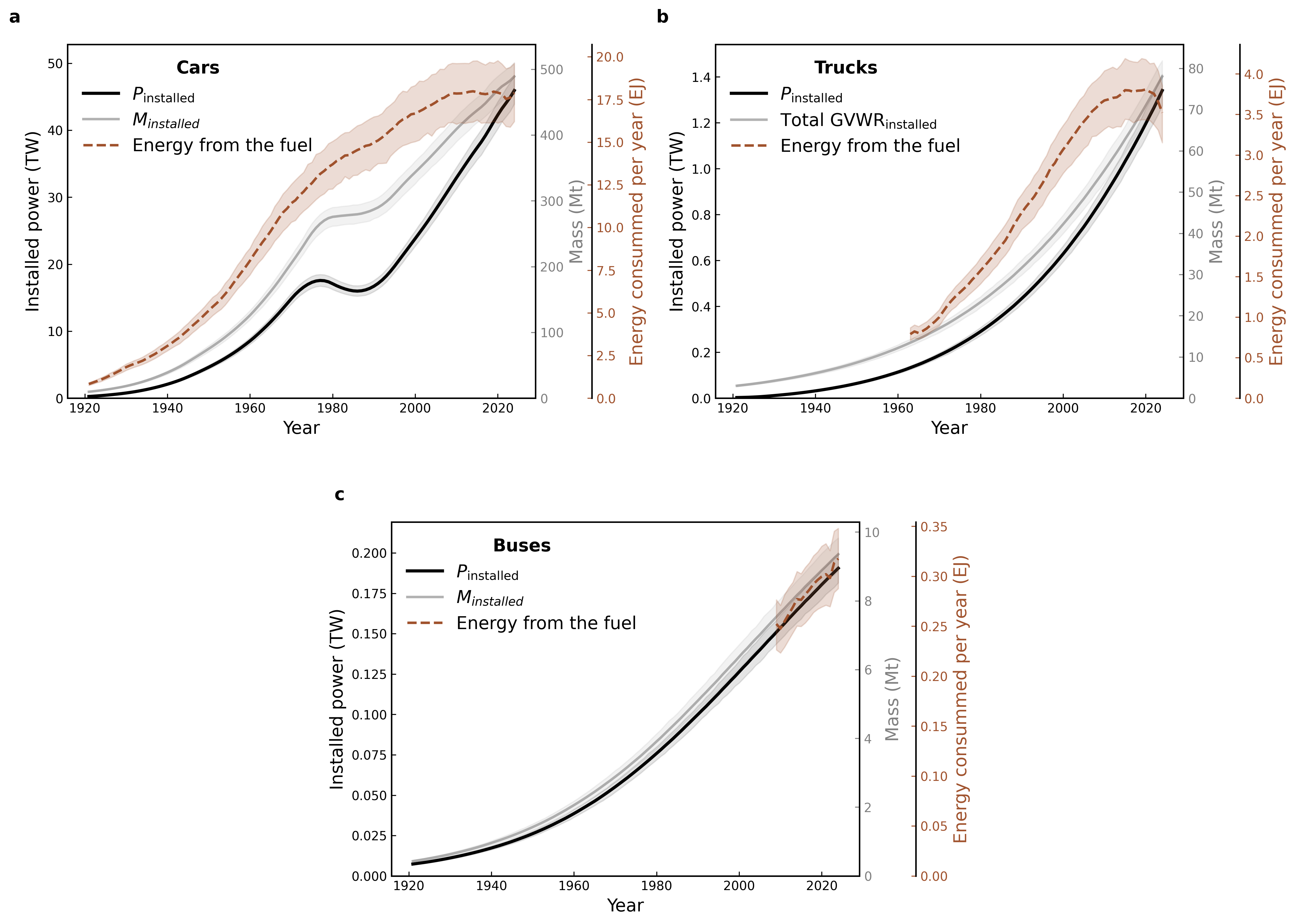}
    \caption{\textbf{Evolution of the total installed motor power (black), mass (grey) and fuel energy consumption (dashed orange) for the three types of vehicles.} \textbf{a}, Cars. \textbf{b}, Trucks. \textbf{c}, Buses. For trucks, the installed mass corresponds to the total Gross Vehicle Weight Rating (GVWR), defined as the maximum weight the tractor unit can carry, including its own curb weight. The shaded areas quantify the uncertainty associated with each curve (95\% confidence interval). }
    \label{fig:reconstruction_ipecs}
\end{figure}

\begin{figure}[h!]
    \centering
    \includegraphics[width=\textwidth]{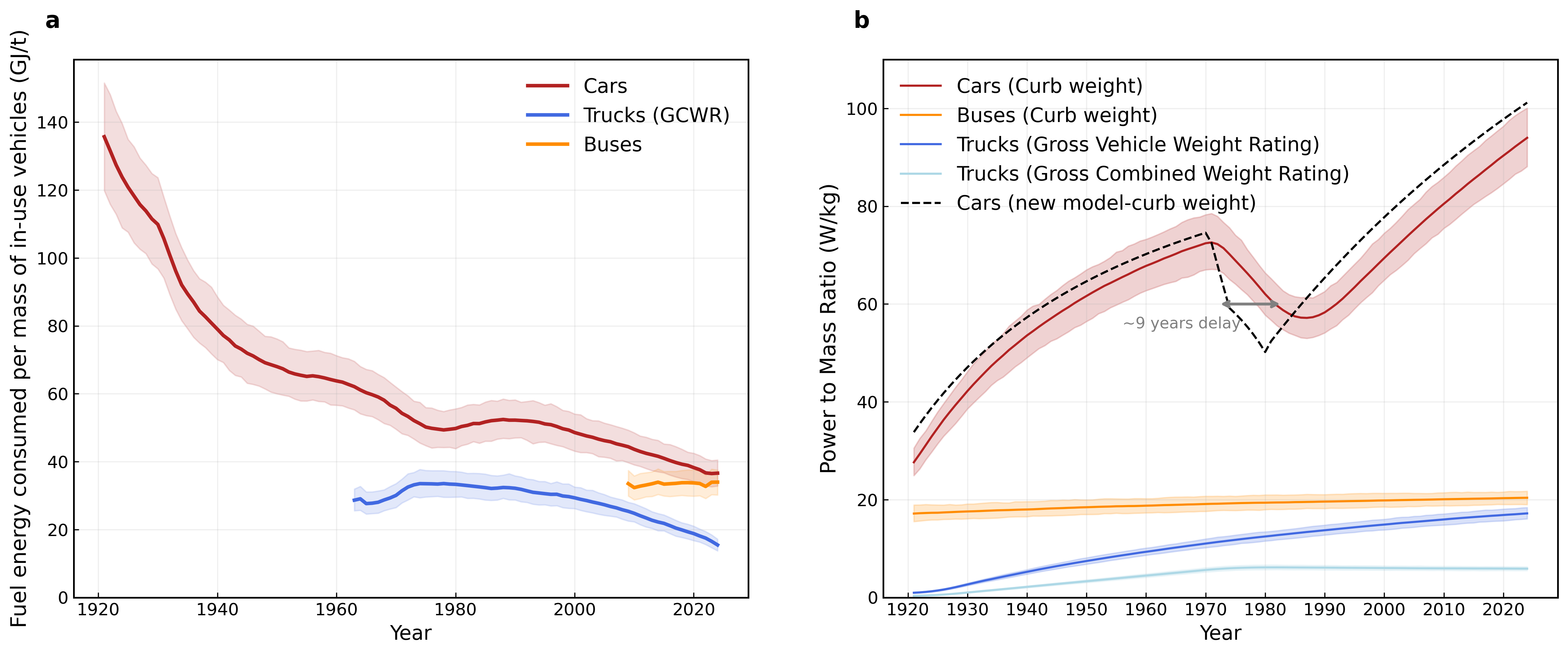}
    \caption{\textbf{Energy consumption and motor power per unit mass of in-use vehicles.} \textbf{a,} Energy intensity of in-use vehicles: the fuel energy used each year per ton of in-use vehicles (for trucks the ton of vehicle correspond to the total Gross Combined Weight Rating GCWR, the maximum weight that the truck can tow by adding trailer in addition to the GVWR). The shaded areas quantify the uncertainty associated with each curve (95\% confidence interval). \textbf{b}, Evolution of the power-to-mass ratio for cars (red), buses (yellow) and trucks. Curb weight is used for cars and buses while Gross Vehicle Weight Rating GVWR is used for full trucks without trailers (blue) and Gross Combined Weight Rating GCWR is used for full trucks with trailers (cyan). The shaded areas quantify the uncertainty associated with each curve (95\% confidence interval). Black dotted line shows the power-to-mass ratio of new car sold. The fleet inertia delay estimated at 9 years is indicated with an arrow. }
    \label{fig:pmr}
\end{figure}

Fig.\,\ref{fig:reconstruction_ipecs} puts the reconstructed total motor power and mass of cars, trucks and buses in perspective with their energy consumption. The rise in power and mass is strong and continuous for all U.S. vehicles -- except for a short-term disruption for cars in the 1970s, after which growth has resumed. For cars and trucks, this growth contrasts with fuel consumption which has plateaued since 2010, see Fig.\,\ref{fig:reconstruction_ipecs}a,b. For buses, however, total motor power, mass and fuel consumption have increased at a similar pace (Fig.\,\ref{fig:reconstruction_ipecs}c) indicating close coevolution between structure and activity. 
The distinct patterns of fuel consumption and motor power per unit mass of in-use vehicles are shown in Fig.\,\ref{fig:pmr}. Unitary fuel consumption is constant for buses, whereas it has declined significantly for cars and trucks (Fig.\,\ref{fig:pmr}a). Despite this decline, the power-to-mass ratio (PMR) -- the actual motor power that is installed to move the mass -- has increased for cars and trucks, whereas it has remained constant for buses (Fig.\,\ref{fig:pmr}b).

For passenger cars, the PMR has quadrupled over the last century, with a notable short-term anomaly between 1973 and 1985. This anomaly stems from the sharp decline in the PMR of newly sold vehicles in 1973, which we examine in section\,\ref{subsec:trajectoire_voitures}. The 9-year lag in the PMR shown in Fig.\,\ref{fig:pmr}b, reflects the time required for the drop in PMR for new vehicle sales to propagate to a comparable order of magnitude in the total in-use vehicle stock.

For trucks, the PMR increases when calculated using the tractor unit’s power relative to its Gross Vehicle Weight Rating (GVWR). However, this trend stagnates when the PMR is assessed against the Gross Combination Weight Rating (GCWR). Comparing these two trends reveals that the added power is necessary to tow significantly heavier trailers. By increasing engine power to accommodate heavier loads, manufacturers have maintained a relatively constant PMR for complete trucks over the past 40 years.
The stable PMR in buses and trucks reflects optimized power requirements tailored to their specific payloads, particularly when contrasted with passenger cars. In cars, the continuous rise in PMR indicates a lack of power optimization, directly tying the oversizing of the fleet to the decline in its usage factor.

\begin{figure}[h!]
    \centering
    \includegraphics[width=\textwidth]{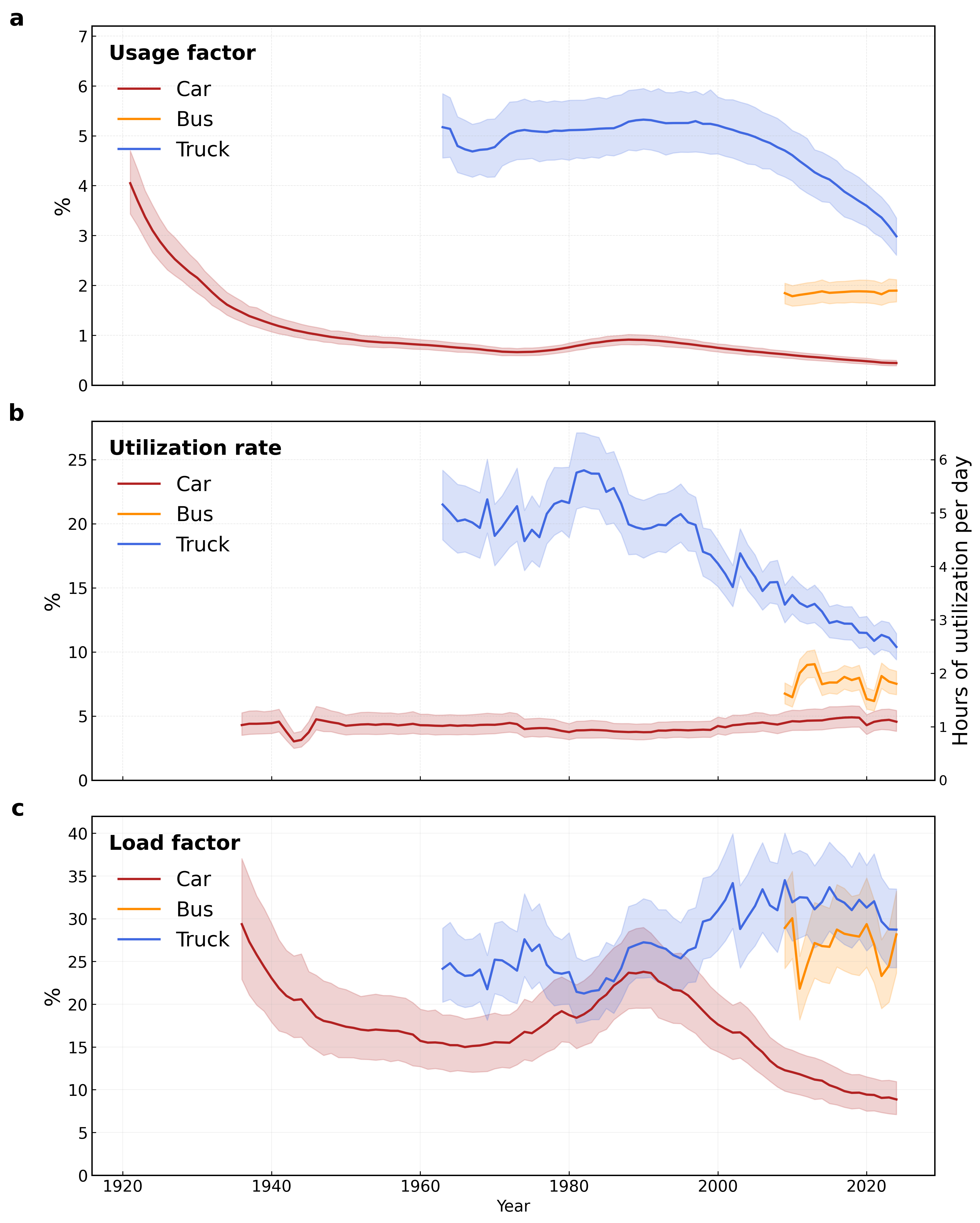}
    \caption{\textbf{Evolution of the usage factor $f$ and it decomposition ratios for cars and buses and trucks.} \textbf{a}, Usage factor: $f = E_{\text{used}}/(P_{\text{installed}} \times T_{\text{year}})$. \textbf{b}, Utilization rate:  $T_{\text{used}}/T_{\text{year}}$ with $T_{\text{used}}$ the duration of usage of the vehicle. \textbf{c}, Load factor: $P_{\text{mean}}/P_{\text{installed}}$ with $P_{\text{mean}}$ the mean power over the duration of the usage and $P_{\text{installed}}$ the power capacity of the vehicle. The decomposition of the usage factor is such that $f = \text{Utilization rate} \times \text{Load factor}$. Shaded areas represent the 95\% confidence intervals.}
    \label{fig:pannel}
\end{figure}

Fig.\,\ref{fig:pannel} highlights the usage factor $f$ per vehicle fleet and its decomposition into a utilization rate (share of time the structure operates, Fig.\,\ref{fig:pannel}b) and a load factor (the fraction of installed power that is effectively used in average during operation, Fig.\,\ref{fig:pannel}c). The decoupling between structure and activity is reflected by the decline of the usage factor for cars (from 2.6\% in 1920 to 0.4\% in 2024) and trucks (from 5.2\% in 2000 to 3\% in 2024). 
The analysis of the underlying patterns shows a long-term stability in the utilization rate for cars (at 4\%, i.e. 1h day$^{-1}$), a relative stability for buses and a sharp decline for trucks since 1980. The load factor shows a value around 25\% for all vehicles until 1990 (Fig.\,\ref{fig:pannel}c) and a shift for cars from this date on to a distinct zone, stemming from engine oversizing over the last decades.


\subsection{Cars oversizing over time}
\label{subsec:trajectoire_voitures}
\begin{figure}[h!]
    \centering
    \includegraphics[width=\textwidth]{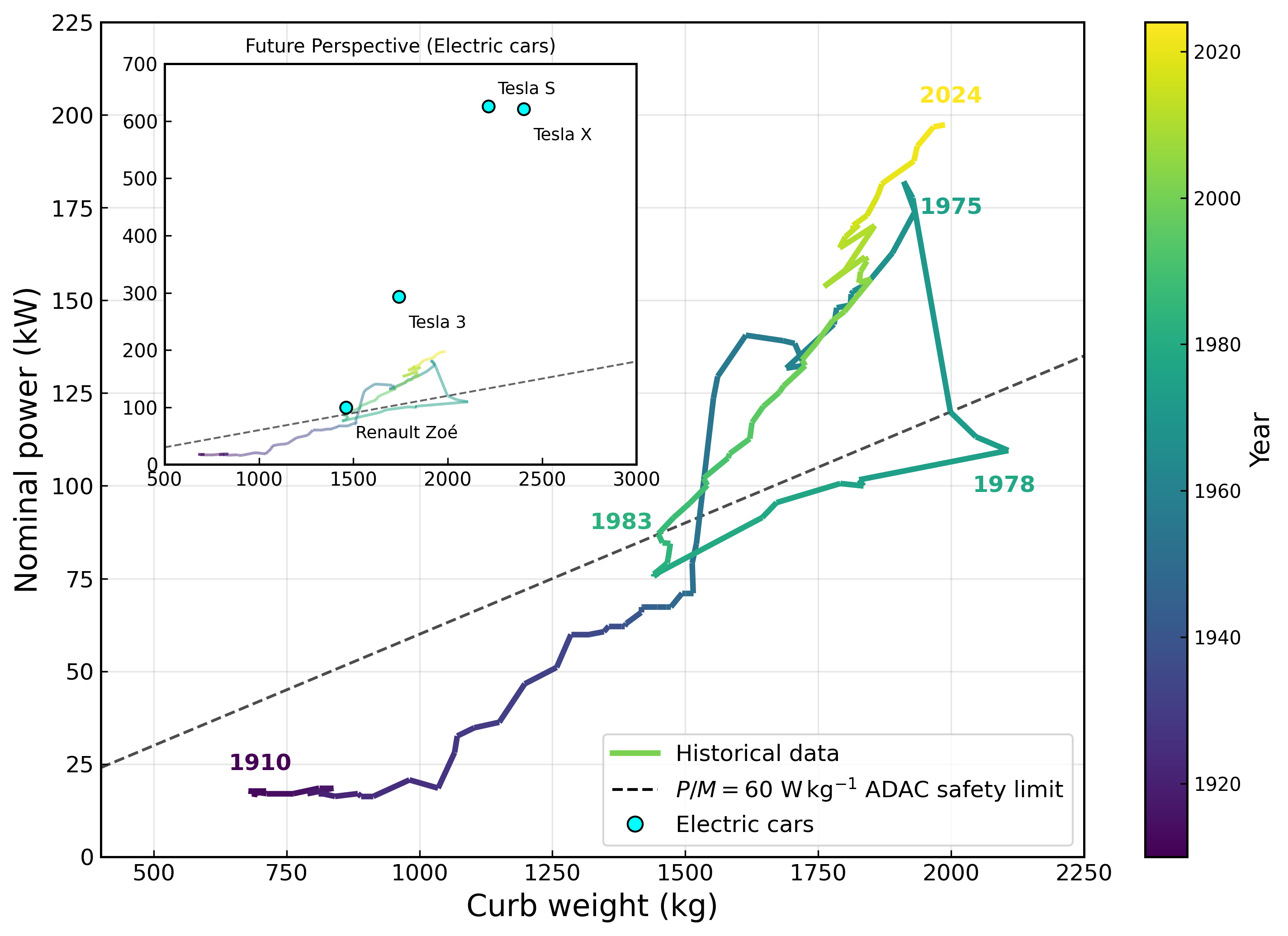}
    \caption{\textbf{Evolution of the average nominal power and curb weight of vehicles sold each year} (color scale). Dotted black line indicates the necessary peak power for a car as stated by the ADAC association. The inset figure shows the characteristics of several models of electric cars. Tesla\,3, S and X are the best-selling models in 2020}
    \label{fig:model_cars}
\end{figure}

The long-term evolution of motor power and vehicle mass for new car models is shown in Fig.\,\ref{fig:model_cars}, with time represented by the color scale. The inset figure situates best-selling electric car models within this historical trajectory. Both power and mass have increased over time, with power rising more sharply than mass. The dotted line represents the peak-effort threshold, defined as the minimum power required for a car to safely merge onto a modern highway, based on ADAC estimates\,\cite{ADAC}. This threshold serves as a reference for PMR requirements of a car. Trajectories with slopes steeper than the dotted line indicate a divergence of car design from actual use requirements.

An anomaly appears in the 1970s trajectory of motor power and mass, where the PMR trajectory forms a distinct loop. In 1975, the power of new cars dropped sharply, followed by a reduction in curb weight between 1978 and 1979. After 1983, both power and mass resumed their upward trend. This temporary disruption was triggered by new EPA regulations\,\cite{Gerard2005} in response to the oil shock. Manufacturers initially adapted by reducing engine power, then later redesigned car structures to lower their weight. From the 1980s onward, engine power resumed increasing faster than curb weight, leading to a steeply ascending PMR slope that surpassed pre-oil shock power and mass levels by the late 2000s.
The characteristics of the best-selling U.S. electric vehicles (EVs) in 2020 (Tesla Models 3, S, and X) are well above the actual thermal car average in both power and mass (inset Fig.\,\ref{fig:model_cars}). By contrast, the Renault Zoé -- a small European electric vehicle -- demonstrates that low-power EVs satisfying ADAC highway safety requirements technically exist, indicating that the trajectory toward heavier, more powerful electric cars reflects market preferences rather than technological constraints.

\newpage
\section{Discussion}

\subsection{Power in the metabolic approach}

The MEFA framework has long provided critical insights into the metabolism of sociotechnical systems by quantifying material stocks and energy flows. Yet, power (watts) has never been systematically incorporated into MEFA, nor explored as a connecting variable between the materials embedded in structures and the energy consumption stemming from activity. This gap limits our understanding of how physical structures enable activity through energy conversion, and whether structure and activity show diverging patterns of evolution depending on the context.

While MEFA has documented declining stock productivity at macro scales\,\cite{Krausmann2017, chen2015}, the distinction between structure and activity has so far received very little scrutiny despite its physical significance. To capture this distinction, we extend the concept of installed power -- traditionally studied through the lens of power generation capacity, never from an energy consumption perspective -- to energy-consuming structures (IPECS). Based on various sources, we reconstruct long-term trends of installed power, mass, and energy use across professional and consumer assets and introduce the usage factor of energy-consuming structures as a generalization of the capacity factor of energy-producing systems\,\cite{IAEA2019}. Applied to the U.S. transportation sector, the usage factor allows us to quantify diverging dynamics between structure and activity, with major differences between professional and consumer assets.

The scarcity of literature on this topic likely reflects the absence of relevant data. While final energy consumption is available by major societal sector at different scales\,\cite{iea2026balances}, consistent long-term time series remain rare, and no equivalent data describe the evolution of the structures -- installed power and mass -- that underpin activity. The century-scale analysis of structure and activity dynamics presented here thus goes far beyond what readily available national and international databases currently support. This work complements previous studies on sociotechnical equipment dynamics over time\,\cite{chen2015}, which focused on the number of in-stock objects, by characterizing the underlying dynamics of equipment with respect to their use.

A key contribution of this work is demonstrating that installed power acts as a connecting variable between material and energy use, with increases in power correlating with increases in mass\,\cite{Caduff2011}. This relationship exhibits dynamics analogous to the rebound effect observed in energy efficiency studies. Just as improvements in energy efficiency can lead to increased total energy consumption by enabling greater activity, gains in PMR -- engines delivering more power per unit mass -- do not necessarily reduce the total mass of vehicles, as they tend to be redirected toward larger, heavier vehicles. A similar mechanism operates on the energy side: increases in motor power can offset efficiency gains, preventing reductions in energy consumption even when the number of passenger-kilometers traveled remains constant. In the specific case of U.S. road transport, total distance traveled has plateaued, yet final energy consumption has not decreased despite motor efficiency improvements. This pattern aligns with the role of motor power in counteracting efficiency gains, though real-world driving conditions (e.g., road smoothness, urban vs. rural mobility shares) also influence vehicle energy efficiency.

We also show that these dynamics differ markedly between professional and consumer assets. Professional vehicles follow optimization dynamics where the usage factor tends to be maximized\,\cite{fred2018trends}, closely aligning activity with structure. Consumer markets, by contrast, exhibit growing overcapacity, where structure increasingly outpaces activity. This contrast underscores the importance of disaggregating analyses by sector and asset type to capture the heterogeneity of structure-activity relationships, and to better target policy interventions aimed at reducing not only energy use but also material immobilization.

\subsection{Decoupling dynamics}
 
Our study reveals a sustained decrease in the usage factor for both cars and trucks, yet these trends are driven by structurally distinct mechanisms: growing redundancy in the truck fleet (with PMR plateauing), reflecting changing vehicle use patterns, versus increasing oversizing in the car fleet. The structure versus activity analysis is relevant for studying diverging structural expansion i.e. when structure growth outpaces activity growth -- that we can qualify as overstructuration.

For professional sectors (here trucks and buses) the usage factor remains more closely tied to operational necessity. The decline in the usage factor since the 2000s observed for trucks is primarily due to a reduction in the distance travelled per vehicle (Extended Data Fig.\,1), which may reflect the growing use of hub-and-spoke logistics\,\cite{doe2024fotw1340}, suggesting increasing redundancy in the fleet.

The decline in the utilization rate of trucks between the 1980s and the 2000s is due to higher speeds (Extended Data Fig.\,2), which is directly compensated by an increase in the load factor owing to higher power use. For buses the usage factor remains stable suggesting optimization of the structure with regards to activity. 

In contrast, the passenger car fleet exhibits a long-term, near-continuous decrease in its usage factor, as the development of the structure has become decoupled from activity, reflected by fuel consumption. The only exception is the period 1975-1983, highlighted in Fig.\,\ref{fig:model_cars}, when the 1975 EPA directive\,\cite{Gerard2005} imposed a reduction in emission (Clean Air Act amendment), that leads to an abrupt and drastic downsizing of engine power, followed by a comparable drop in vehicle mass over subsequent years.
Indeed, the decreasing trend in usage is not due to a decline in energy service but is instead driven by a massive increase in the nominal power and  mass of vehicles. This decoupling marks a key distinction between transport sectors: professional structures remain optimized for utility, whereas personal vehicles have evolved toward significant overcapacity.
The decline in the car usage factor is primarily due to the widening gap between maximal power capacity and actual maximal power requirements, while mean power has remained steady over time (Extended Data Fig.\,3). This gap indicates increasing oversizing, meaning that an ever-growing share of the fleet’s mechanical potential remains unused.

These diverging mechanisms reflect fundamentally different economic logics between professional and consumer assets. For professional uses there is no economic incentive to oversize engines, as this would increase costs without operational benefits. Instead, competition drives the minimization of capital costs per unit production. As a result, nominal power is calibrated to closely match real-world peak power needs, and less excess mass is immobilized in these structures.
This is quantified by the similarity in load factor values for trucks and buses, which are consistently between \,25\% and 35\% when accounting for uncertainty (Fig.\,\ref{fig:pannel}c). Conversely, in the consumer market, this investment logic is absent. Purchasing decisions are instead driven by factors beyond operational requirements, such as performance\,\cite{Ewing2000, MANDYS2021} and social status\,\cite{fujita2022, BRONNMANN2025}, which explains the divergence between growing installed capacity and actual usage intensity.

Moreover, the PMR of cars is rising alongside installed mass, highlighting increasing immobilization of materials. There is no evidence of motor power increasing without a corresponding increase in vehicle mass\,\cite{Caduff2011}. Furthermore, even in a scenario where technology could increase PMR while keeping vehicle mass constant, fuel efficiency would still be negatively affected, as a more powerful engine consumes more fuel\,\cite{Caduff2011, Horsepower_mpg}. This addresses a gap in the material productivity literature\,\cite{Krausmann2017, Schandl2024}, where the distinction between sectors is barely addressed. 
\newline

Beyond market considerations, the physical nature of road transport imposes structural constraints on how intensively a vehicle can be used.
Load factors are structurally low in road transport because vehicles must be capable of delivering a wide range of power output. A driving cycle consists of successive phases of acceleration, cruise speed, and stops, each requiring vastly different power levels\,\cite{Agarwal2021}. Consequently, mean power over a usage period is inherently lower than both the necessary peak power and the nominal power. Nevertheless, by calibrating nominal power closer to peak power needs, manufacturers can help to maximize the load factor.
Furthermore, the duration of utilization can differ by several hours per day depending on whether use is professional or private. Time allocated to car use results from socioeconomic dynamics that have constrained it to approximately one hour per day throughout the past century\,\cite{WANG2024, Shoup2021}. For buses, the temporal intensity ratio is significantly higher besides being pulled down by the presence of school buses in fleet, which operate over shorter durations than public transit buses.

To maintain activity levels while improving sustainability, the amount of structure per unit activity should be minimized. This creates a trade-off between material use and service provision. Some redundancy nonetheless remains necessary to ensure service robustness -- requiring standby vehicles\,\cite{NAP2014, Fleurent2017, JANSUWAN2021}. Mutualization of passenger cars would increase the utilization rate of each unit and may reduce the number of vehicles in stock\,\cite{Keith2024}.

\subsection{Policy implications for sustainability}

Our findings highlight increasing decoupling between structure and activity in consumer assets, resulting in decreasing material efficiency compared to professional sectors. Following the dynamics of professional sectors and aiming for usage factors similar to those of professional assets could help reduce the material use intensity of structures.

The challenge of structural overcapacity takes on new urgency in the context of fleet electrification. The ongoing electrification of the U.S. car fleet\,\cite{Muratori2025} will involve a large-scale replacement of the vehicle fleet and a surge in material demand. Yet, we show evidence that electrification is likely to amplify vehicle oversizing, as best-selling electric models tend to be significantly more powerful and heavier than their combustion-engine counterparts\,\cite{EPA2024trends}. Consequently, with electric vehicles, the gap between nominal power and maximum required power becomes even more pronounced. We demonstrate that to limit mass, power must be limited. Each increase in power is directly correlated with an increase in mass, and this trend suggests a further divergence between installed power and future material requirements for fleet electrification.
 
Ultimately, this poses a critical challenge for the energy transition. The immobilization of critical materials—such as copper, lithium, cobalt, and nickel\,\cite{Olivetti2017}—in oversized energy-consuming structures represents a clear misallocation of resources, particularly as these materials are essential for renewable energy systems and their demand is projected to rise sharply\,\cite{Woodley2024}. This reveals a contradiction in sustainability objectives, as critical materials are being locked into underutilized structures, highlighting untapped potential for resource savings if structure per unit of activity were minimized.

Addressing this misallocation requires targeted policy interventions to directly tackle over-structure. Previous research has focused on reducing vehicle weight through alternative materials\,\cite{tempelman2019, DAS2016}. However, vehicle weight has continued to increase steadily even after the introduction of low-density materials such as plastics and aluminum to replace steel\,\cite{EPA2024trends}, and -- as we demonstrate -- it scales with power.

Since engine output power determines a vehicle's mechanical capacity, limiting power would reduce the mass of materials immobilized in the fleet\,\cite{Caduff2011} without compromising usage levels. This could be achieved through policies capping the power of newly sold vehicles, similar to the intensity ratio adjustments implemented after 1973\,\cite{Gerard2005}.

Additionally, as maximum required power is partly determined by highway speed limits, reducing speed limits would lower the minimum power needed for safe driving, providing another lever to curb oversizing and material demand. Both approaches have proven effective. The 1975 EPA regulations (Fig.\,\ref{fig:model_cars}) and Japan's Kei-car program\,\cite{LIPSCY2013} -- which imposed motor power and speed limits in exchange for tax incentives -- demonstrate that such policies can prevent vehicle oversizing and reduce material demand.

Finally, while vehicle sizing should reflect actual usage, motor power is currently decoupled from real-world needs, as peak required power is typically far lower than nominal motor power. Public communication campaigns could help shift consumer preferences away from over-structured vehicles toward designs aligned with actual mobility needs.

Future research should extend installed power reconstruction to other energy-consuming structures (e.g., in the residential sector) to assess potential decoupling between final energy consumption and installed power. Quantifying material savings from power regulation remains an open question and a priority for future work.

\section{Material and methods}

\subsection{Theory elements}

The installed power of energy consuming structures (IPECS) is defined as the aggregate nominal power of a given fleet of energy consuming structures. It extends the concept of installed capacity, conventionally applied to energy producing structures, to energy consuming structures.
Using IPECS, we further define the usage factor $f$ of energy consuming structures as the ratio between the energy effectively consumed or used ($E_{\text{used}}$), and the theoretical energy that would be consumed if the installed power ($P_{\text{installed}}$) were used continuously at full nominal power over one year. Its calculation follows the same principle as the capacity factor commonly used for energy production plants~\cite{IAEA2019}, but applied to energy consuming structures:
\begin{equation}
    f = \frac{E_{\text{used}}}{P_{\text{installed}} \times T_{\text{year}}}
     \label{eq:usage_factor}
\end{equation}
We then decompose the usage factor, following a Kaya-like approach~\cite{Kaya}, into the product of the load factor and the utilization rate.
\begin{equation}
    f = \frac{E_{\text{used}}}{P_{\text{installed}} \times T_{\text{year}}} =  \frac{P_{\text{mean}}}{P_{\text{installed}}} \times \frac{T_{\text{used}}}{T_{\text{year}}}
     \label{eq:kaya}
\end{equation}
where $P_{\text{mean}}$ denotes the mean power during the period of use $T_{\text{used}}$ of the structure. This decomposition can be further extended by breaking down the load factor into two components.
\begin{equation}
 f = \frac{P_{\text{mean}}}{P_{\text{peak}}} \times \frac{P_{\text{peak}}}{P_{\text{installed}}} \times \frac{T_{\text{used}}}{T_{\text{year}}}
\end{equation}
where $P_{\text{peak}}$ denotes the maximum power required by the structure during its period of use.
The ratio $P_{\text{mean}}/P_{\text{peak}}$ characterizes the power-demand profile associated with the activity performed by the structure during a period of use. More specifically, it captures the gap between the peak power required during use and the mean power effectively demanded over the same period. A low value therefore indicates a highly peaked power profile, in which short periods of high power demand coexist with much lower average power requirements.
The ratio $P_{\text{peak}}/P_{\text{installed}}$ indicates the degree of power oversizing, that is, the extent to which the installed nominal power exceeds the maximum power actually required during use. This ratio is mainly determined by the design and sizing choices made by the manufacturer.
Finally, the utilization rate $T_{\text{used}}/T_{\text{year}}$ represents the share of time during which the structure is used over the year. This ratio decreases when the number of redundant structures per user increases.
Analyzing each of these ratios individually helps disentangle the different mechanisms that shape the usage factor of a given structure (e.g. redundancy, oversizing, etc.).

\subsection{Reconstruction of in-use vehicles stocks evolution}

To simulate vehicle retirement over time, we use a Weibull survival function:
\begin{equation}
W_{i,j} = \exp \left[- \left( \frac{i-j}{\lambda(j)} \right)^k \right],
\end{equation}
where $W_{i,j}$ denotes the probability that a vehicle sold in year $j$ remains in the stock in year $i$, $\lambda(j)$ is the scale parameter associated with the sales cohort $j$, and $k$ is the shape parameter of the Weibull distribution.

Following the literature~\cite{Oguchi2015}, the Weibull shape parameter $k$ is set to $3.6$. To estimate the time evolution of the scale parameter $\lambda$, we fit a logistic function to $\lambda(j)$ using the following balance equation:
\begin{equation}
N_{\text{cars}}(i) = \sum_{j=0}^{i} W_{i,j} \cdot S_{\text{cars}}(j),
\end{equation}
where $N_{\text{cars}}(i)$ denotes the number of cars in the stock at the end of year $i$, and $S_{\text{cars}}(j)$ denotes the number of cars sold during year $j$. We allow $\lambda$ to vary over time because it is related to the mean vehicle lifetime through
$\lambda = \frac{\text{Lifetime}}{\Gamma\left(1 + 1/k\right)}$,
and vehicle lifetimes have increased over time according to the FHWA report~\cite{usdot2002ournationshighways}. Further details are provided in the Supplementary Materials.

Because sales data for trucks and buses are not available over a sufficiently long period, sales are reconstructed from stock data as follows:
\begin{equation}
    S_{\,v} = W^{-1} \cdot N_{\,v}  \quad \text{where  } v \in \{\text{bus, truck}\}
\end{equation} 
where $W$ is the lower triangular survival matrix, with elements $W_{i,j}$ for $j \leq i$ and 0 otherwise.

Since the lifetime assumptions implied by regulations for public buses~\cite{bus_lifetime} and trucks~\cite{truck_lifetime} are close to the lifetimes estimated for cars, we use the $\lambda$ values calibrated for cars and apply them to buses and trucks. Further details on the stock reconstruction procedure are provided in the Supplementary Materials.

\subsection{Installed motor power, vehicle mass, and power-to-mass ratio by vehicle type over time}

For each vehicle type, we compute:
\begin{align}
     P_{\,v,\text{installed}} (i) &= \sum_{j=0}^{i} W_{i,j} \cdot p^{\text{sold}}_{\,v,\text{ nominal}}(j) \cdot S_{\,v}(j) \quad \text{where  } v \in \{\text{car, bus, truck}\} \\
     M_{\,v} (i) &= \sum_{j=0}^{i} W_{i,j} \cdot m^{\text{sold}}_{\,v}(j) \cdot S_{\,v}(j) \\
     \text{PMR}_{\,v} (i) &= P_{\,v,\text{installed}}/M_{\,v} (i) 
\end{align} 
where $P_{\,v,\text{installed}}(i)$, $M_{\,v}(i)$ and $\text{PMR}_{\,v}(i)$ denote, respectively, the total installed motor power of the fleet, the total mass of the fleet, and the fleet-level power-to-mass ratio for vehicle type $v$ in year $i$.

The nominal power of newly sold cars, $p^{\text{sold}}_{\text{nominal}}$, is obtained from the EPA Data Explorer for the period 1975--2024. For earlier years, we combine several historical sources to reconstruct fleet characteristics back to 1910, based on the three best-selling car models in each year. Further details are provided in the Supplementary Materials.

For trucks and buses, because aggregated historical data were not available, we adopted a representative-vehicle approach. For each decade, a popular or characteristic model was selected as a proxy for the technological standard of that period. The nominal power of these representative vehicles, as well as their mass in the case of buses, was collected and used as a set of reference points. A linear regression was then fitted to these decade-level data points to reconstruct the corresponding historical trends. Further details are provided in the Supplementary Materials.

A useful metric for assessing vehicle performance is the power-to-mass ratio (PMR). In 2024, the ADAC association identified a minimum PMR of $60$\,W/kg for passenger cars as necessary to ensure safe highway driving~\cite{ADAC}.

For cars and buses, the mass used to compute the PMR corresponds to curb weight. For trucks, we use the Gross Vehicle Weight Rating (GVWR), which corresponds to the curb weight of the truck plus its maximum allowable payload, excluding trailers. We also consider the Gross Combined Weight Rating (GCWR), which corresponds to the GVWR plus the maximum allowable trailer weight.

\subsection{Usage factor of road transport vehicles}

When applied to road transport, the usage factor of vehicle type $v \in \{\text{car, bus, truck}\} $ can be written as: 
\begin{equation}
f_v(t) = \frac{E_{\,v,\text{ fuel}}(t)\times \eta_{\,\text{bte}}(t)}{P_{\,v,\text{installed}}(t) \times T_{\text{year}}}
\end{equation}
with $\eta_{\,\text{bte}}$ denotes the brake thermal efficiency of the engine.

In this study, we consider only fuel energy consumption, as the share of electric vehicles in the stock remains negligible over the period considered~\cite{Fadhil2025,Isenstadt2025}. Extending the analysis beyond 2024 would require accounting for the electricity consumed by electric vehicles.

To decompose the usage factor, we compute the following two quantities over time:
\begin{align}
    T_{\text{used,v}}(t) &= \frac{d_v(t)}{a_v(t)} \\
    P_{\text{mean},v}(t) &= 
                \frac{E_{\text{fuel},v}(t) 
                \times
                \eta_{\text{bte}}(t)}{T_{\text{used},v}(t)}
\end{align}
Here, $d_v(t)$ denotes the annual distance travelled by vehicle type $v$, and $a_v(t)$ denotes its average speed (Extended Data Fig.\,4). These data are provided in the Supplementary Material.

We then compute the ratios $T_{\text{used},v}/T_{\text{year}}$ and $P_{\text{mean},v}/P_{\text{nominal},v}$, where the average nominal power per vehicle is defined as: $P_{\text{nominal},v} = P_{\text{installed},v}/N_v$

\subsection{Data collection}

In this study, the category \textit{car} encompasses all light-duty vehicles, including SUVs, pickups, and vehicles for both private and professional use (e.g., taxis). \textit{Truck} refers specifically to heavy-duty trailer trucks, while \textit{Bus} accounts for all types of bus services, including school, city transit, intercity, and airport shuttles.

Data of vehicles registrations, fuel consumption, miles travelled are collected for the Bureau of Transports statistics and the FHWA Data Explorer and then aggregated as described in the Supplementary Material. Fuel consumption data were smoothed using a centered moving average, with an 11-year window whose size was reduced near the edges to preserve the full time period.

Car sales data were obtained from Ward’s Automotive Group\,\cite{WardsAutomotive2010} and Statista\,\cite{StatistaUSVehicleSales2025}.

The evolution of brake thermal efficiency over time was compiled from the literature\,\cite{Yousif2022,GHANIM2026} and fitted using a four-parameter logistic curve, as detailed in the Supplementary Material.

Average vehicle speed data were obtained from federal reports\,\cite{2017nhts,Raff1951}, as described in the Supplementary Material.

To convert fuel volumes reported by the FHWA from gallons to energy units, we use an energy density of $130$ MJ/gal\,\cite{MAZLOOMI2012}.

Primary and reconstructed data are available in an Excel spreadsheet in Supplementary Data.

\subsection{Uncertainty calculation}

Uncertainties were propagated using Monte Carlo simulations ($n = 1000$). For primary data, including fuel consumption and total vehicle-miles travelled, an uncertainty of 5\% was assumed. For in-use vehicle stocks, a 5\% uncertainty was applied to state-level data and then propagated to the national total. For car production, Poisson uncertainty was assumed.

A 10\% uncertainty was assigned to the Weibull shape parameter $k$. This uncertainty was then propagated to the estimation of the Weibull scale parameter $\lambda$. For average vehicle speeds, the fitting error was used for trucks, while an uncertainty of $2$ miles per hour was applied to cars and buses. For brake thermal efficiency, the fitting error was used. Finally, for the nominal power and mass of newly sold vehicle models, we assumed a 10\% uncertainty, to account for the variability associated with the decade-level reconstruction of vehicle characteristics.

\clearpage
\section{References}
\bibliographystyle{naturemag}
\bibliography{references}   

\newpage
\section{Extended data}

\renewcommand{\figurename}{Extended Data Figure}
\setcounter{figure}{0}
\begin{figure}[h!]
    \centering
    \includegraphics[width=\textwidth]{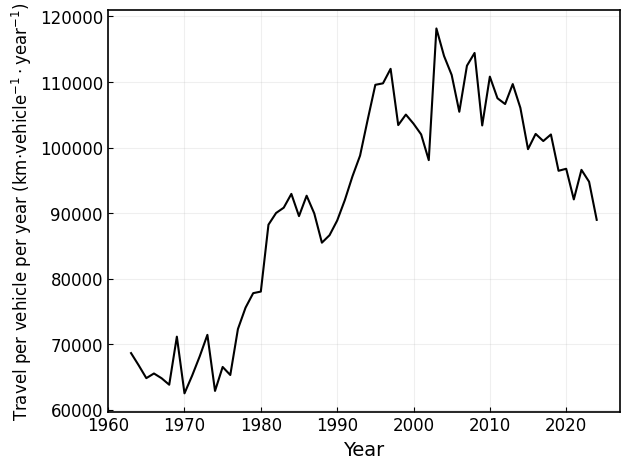}
    \caption{\textbf{Average travel per combination truck per year (kilometer).}}
\end{figure}

\clearpage

\begin{figure}[h!]
    \centering
    \includegraphics[width=\textwidth]{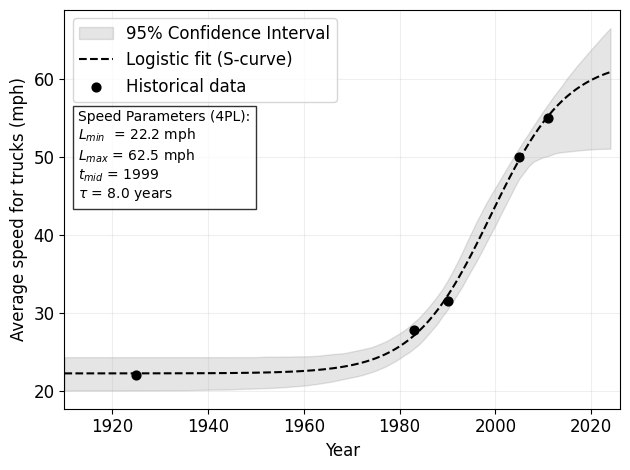}
    \caption{\textbf{Average speed of trucks (miles per hour).} A four-parameter logistic model was fitted to the historical data. The shaded area represents the 95\% confidence interval.}
\end{figure}

\clearpage

\begin{figure}[h!]
    \centering
    \includegraphics[width=\textwidth]{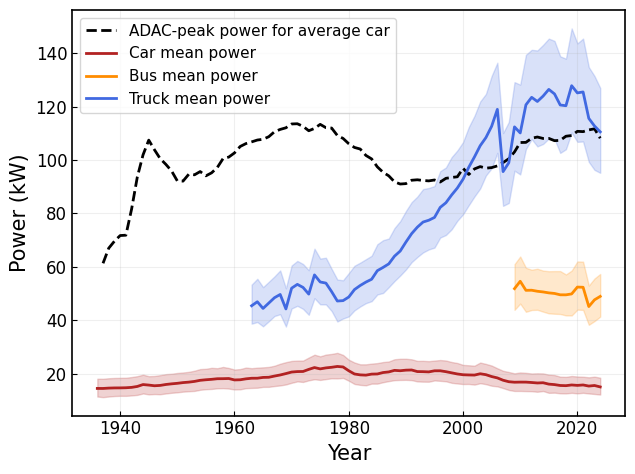}
    \caption{\textbf{Mean power during use for the three vehicle types (kW). The dotted line indicates the peak power required according to the ADAC criterion.} The required peak power is calculated using the ADAC threshold of $60$ W.kg$^{-1}$ and the mass of the average car.}
\end{figure}
\clearpage

\begin{figure}[h!]
    \centering
    \includegraphics[width=\textwidth]{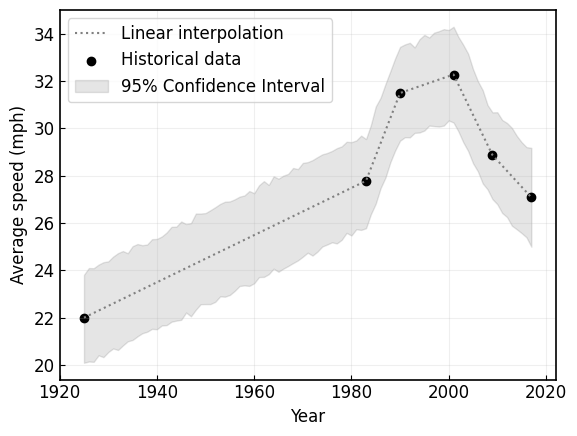}
    \caption{\textbf{Average speed of cars and buses (miles per hour).} The shaded area represents the 95\% confidence interval.}
\end{figure}

\end{document}